\documentclass[12pt,preprint]{aastex}
\usepackage{natbib}
\bibpunct{(}{)}{;}{a}{}{,}

\shorttitle{Origin of lithium enrichment in K giants}
\shortauthors{Bharat et al.}

\begin{document}
\title{Origin of lithium enrichment in K giants}
\author{Yerra Bharat Kumar$^{1}$, Bacham E. Reddy$^{1}$, David L.
Lambert$^{2}$ }
\affil{$^{1}$ Indian Institute of Astrophysics, Bengaluru, India-560034}
\affil{$^{2}$ McDonald Observatory, The University of Texas, Austin,
Texas, USA 78712}

\begin{abstract}
In this {\it Letter}, we report on a low-resolution spectroscopic survey for
Li-rich K giants among 2000 low mass ($M\leq 3M_\odot$) giants spanning the
luminosity range from below to above the luminosity of the clump.
Fifteen new Li-rich giants including four super Li-rich K giants
($\log\epsilon$(Li) $\geq 3.2$) were discovered.
A significant
finding is that there is a  concentration of
Li-rich K giants at the luminosity of the clump or red horizontal branch.
This new finding is partly a consequence of the fact that our low-resolution
survey is the first large survey to include giants well below and above the RGB bump and clump
locations in the HR diagram. 
Origin of the lithium enrichment may be plausibly attributed to the
conversion of $^3$He via $^7$Be to $^7$Li by the Cameron-Fowler mechanism
but the location for  onset of the conversion is uncertain. Two possible
opportunities to effect this conversion are discussed:
the bump in the first ascent of the red giant branch (RGB) and the He-core
flash at the
tip of the RGB.
The finite luminosity spread of the Li-rich giants serves to reject
the idea that Li enhancement is, in general, a consequence
of a giant swallowing a large planet. 
\end{abstract}


\section{Introduction}

Lithium is readily destroyed within a main sequence star  and may survive
only
in a very thin
surface layer provided that mixing between this layer and the interior
does not
occur. Surviving surface Li is greatly diluted as a star evolves to
become a red giant
because an extensive convective envelope develops.
Standard stellar evolutionary models predict Li dilution by a factor of about
60 for low-mass solar-metallicity stars \citep{iben1967a,iben1967b}.
Thus, a star of approximately solar metallicity
 with an initial (say, meteoritic) Li abundance of
$\log\epsilon$(Li) $\simeq$
3.2 is predicted to have a Li abundance of 1.5 on the red
giant branch (RGB).
In general, observations of Li abundances in K giants
confirm these predictions in the sense that the predicted Li abundance is
the upper
limit to observed Li abundances except for  extremely
rare cases exhibiting a higher Li
abundance.
For many RGB stars, Li is less than the predicted
\citep{brown1989,mallik1999},
and this is likely due to a combination of enhanced destruction of Li by
mixing within a giant's
main sequence progenitor and continued destruction of Li in a giant.

This comfortable picture of destruction and dilution
was demonstrated to be incomplete by the serendipitous discovery of
an unusually large amount of Li in a K giant \citep{wallerstein1982}.
Since then
a few more Li-rich K giants of solar metallicity
 have been discovered  including examples
with Li
abundances
at or even exceeding the expected
 maximum Li abundance of  main sequence
stars.
The first systematic search for Li-rich K giants
\citep{brown1989}
demonstrated the rarity of
Li-rich giants, just about 1$\%$, in the solar neighborhood.

In this {\it Letter}, we present results from a large survey of K giants
designed to
find additional
Li-rich examples, locate their appearance in the HR diagram and so provide
essential material for testing hypotheses about Li-enrichment.
For example,
Li production may be associated with the bump in the luminosity function for
stars on the RGB
as suggested
from observations by \citet{charbonnel2000}. Depending on mass and composition the bump occures
on RGB between luminosities, log (L/L$_{\odot}$) = 1.5 - 2.2, 
and this  has been  predicted \citep{iben1968} as a region where evolution of a giant is slowed down
creating a bump in the luminosity function of stars along the RGB. Internally, the bump is
associated with the H-burning shell crossing over of H-discontinuity.
Additionally or alternatively, the He-core flash may be the trigger for
Li production and the creation of Li-rich red clump (horizontal
branch) stars. Discrimination between bump and clump is possible 
because the two groups appear in somewhat different places in the
HR diagram. 
Our  2000 giants provide a
reference sample
comfortably spanning the luminosity of the clump and the bump.
Also, the survey designed to test the suggestion of planet engulfment
scenario by including giants from all the way from the base of RGB to well above the clump/bump luminosities.
With Brown et al.'s survey, a total of 2644 giants covering 
a wide range of luminosities and effective temperatures represent
a potent database in which
to search for clues to the origin of the Li enrichment in
K giants.

This {\it Letter} is
restricted to discussing the location of the Li-rich stars in the
Hertzsprung-Russell diagram.
Full details of
the survey will be provided in a subsequent publication.

\vskip 0.1cm

\section{Sample Selection and Observations}

K giants with accurate astrometry (parallaxes and proper motions) were chosen
from the Hipparcos catalogue \citep{perryman1997,vanleeuwen2007} to span
the luminosity bump on the RGB. Stars were selected according to their
(B-V) color
such that stars were confined to the mass range of 0.8 - 3.0$M_{\odot}$.
Filters were applied to select
nearby (d $\leq$ 150 pc) bright  (m$_{V}$ $\leq$ 8) stars  within
the declination range of $-$60 $\geq$ $\delta$ $\leq$ 80 deg.
These criteria
resulted in a total of 2000 K giants.
All of which were observed at low spectral
resolution (R=$\lambda$/$\delta \lambda$ $\leq$ 3500)   at one of three
telescopes:
the 2-m Himalayan Chandra Telescope (HCT) with
the faint object spectrograph camera (HFOSC)
at the Indian Astronomical
Observatory (IAO), Hanle;  the  1-m Zeiss telescope fitted with the
universal
astro-grating spectrograph (UAGS), and the 2.34-m Vainu
Bappu Telescope (VBT) fitted with the optometrics medium resolution
spectrograph (OMRS)
at the  Vainu Bappu Observatory (VBO), Kavalur.

Sample low resolution spectra showing Li and Ca features is shown in
Fig.~1. 
The K giants
with a moderate to strong Li doublet at 6707.7 \AA\ were identified and
pursued at
high resolution
(R=60,000) spectroscopy with the 2.34-m VBT equipped with
the echelle spectrograph \citep{rao2005} and/or with the
2.7-m Harlan J. Smith telescope with Robert G. Tull
coud\'{e} echelle spectrometer \citep{tull1995} at the McDonald Observatory.
The raw data were processed under standard tasks in the $IRAF$
package.\footnote{The IRAF software is distributed by the National Optical
Astronomy Observatories under contract with the National Science Foundation.}

The  parameters luminosity and effective temperature are
obtained in identical
fashion for the stars in our and Brown et al.'s survey.
Luminosities are obtained using parallaxes from the
Hipparcos catalogue and the magnitudes (m$_{V}$) taken from the SIMBAD
data base. Effective
temperatures ($T_{\rm eff}$)
are derived using the (B-V)
color and the calibration given in \citet{alonso1999}.
Since sample stars are nearby, the effects of interstellar extinction are
not taken
into account. However, we note that by neglecting reddening
we would have underestimated $T_{\rm eff}$
by about 50 - 100K considering canonical reddening of about E(B-V)
$\approx$ 0.03 - 0.05
for giants of distances of 100 to 200 pc. Neglecting reddening
will not affect the $T_{\rm eff}$ of  Li-rich K giants because these
temperatures were derived using iron lines in the  high resolution
spectra. The temperatures derived from high resolution spectra
are in good agreement (within $\pm$100K)  
with the values derived for the same stars using the (B-V)
color and the \citet{alonso1999} calibration.
\begin{figure}[ht]
\epsscale{0.8}
\plotone{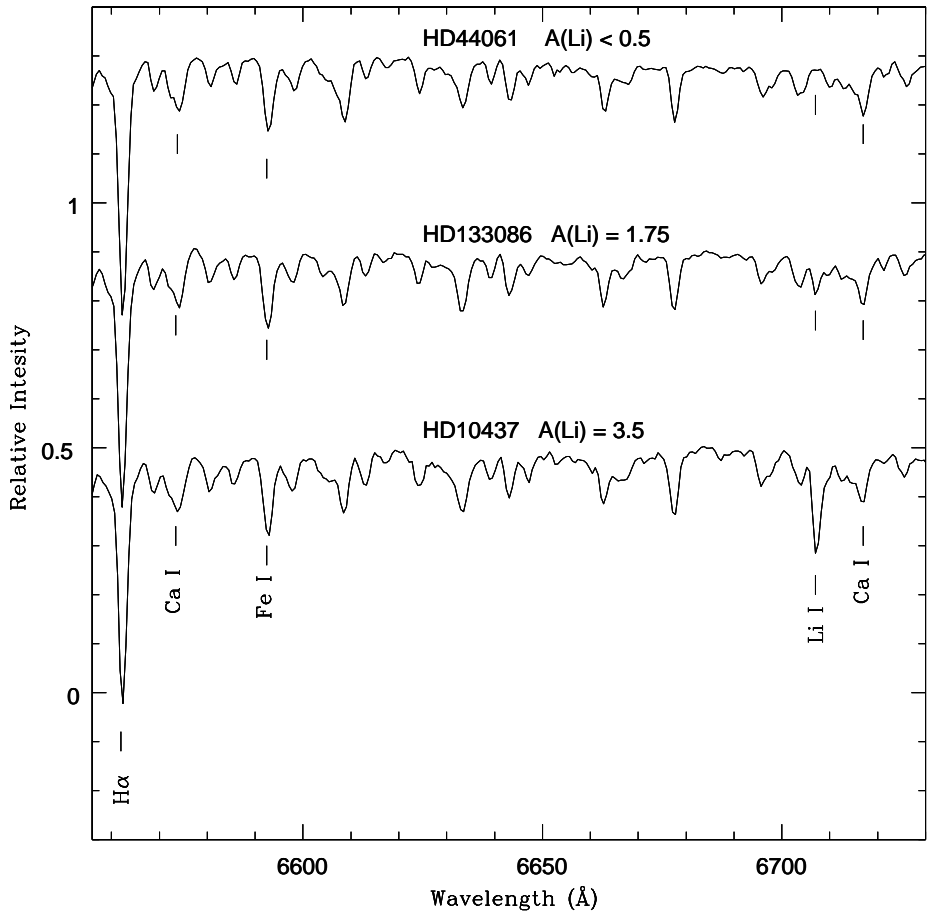}
\caption{Sample low resolution spectra of three K giants. The Li line at 6707\AA\ and Ca at 6717\AA\
which are used in the calibration are shown.
\label{fig1}}
\end{figure}


\section{Lithium Abundances}

Lithium abundances of the stars observed at low resolution were estimated
from the
strength of the Li\,{\sc i}  6707.7 \AA\ doublet relative to that of
  the  Ca\,{\sc i} line at
6717.6 \AA.
An empirical relation between the line-depth ratio of Li and Ca lines and the
Li abundance
was derived from known Li-rich K giants with Li abundances from
high-resolution
spectra. 
Fifteen  K giants in our survey
were estimated to be Li-rich on the basis of the low
resolution spectra.
Accurate abundances for these stars were then obtained from
high resolution spectroscopic analysis (Table~1).
Previously known Li-rich giants of which several were reobserved and
 reanalysed are listed
in Table~2.
Our survey confirms the earlier survey \citep{brown1989}
that the Li-rich K giants are rare.

Tables~1 and 2 give
the  atmospheric parameters ($T_{\rm eff}$, $\log g$, and [Fe/H])
derived from
iron lines in the high-resolution spectra. A majority of the entries for
Table~2 are taken from the papers reporting the Li abundances.  The LTE
and Non-LTE Li
abundances are tabulated where the latter are based on the recipe given
in \citet{lind2009}. In the Tables, we also present $^{12}$C/$^{13}$C
values for the Li-rich K giants. The values of
$^{12}$C/$^{13}$C measurements
are based on molecular lines in the 8003\AA\ region using a similar procedure
to that described in \citet{bharat2009}. For a few giants carbon ratios couldn't be
determined owing to their large rotation and as a result broad spectral features. 

\section{The Hertzsprung-Russell Diagram}

Giants from our and Brown et al.'s  surveys are shown on the HR diagram in
Fig.~2 along
with
evolutionary tracks \citep{bertelli2008} computed
for solar metallicity, [Fe/H] = 0.0, and stellar masses ranging from
0.8$M_{\odot}$
to 3$M_{\odot}$.
All stars observed only at low resolution are assumed to have a solar
metallicity. For Li-rich K giants actual metallicities measured from the
high resolution
spectra are adopted and most are close to solar values. Errors
derived from
the quoted uncertainties in the parallaxes and in the derivation of
temperatures for the Li-rich K giants are marked in Fig.~2.

\begin{figure}[ht]
\epsscale{1.2}
\plottwo{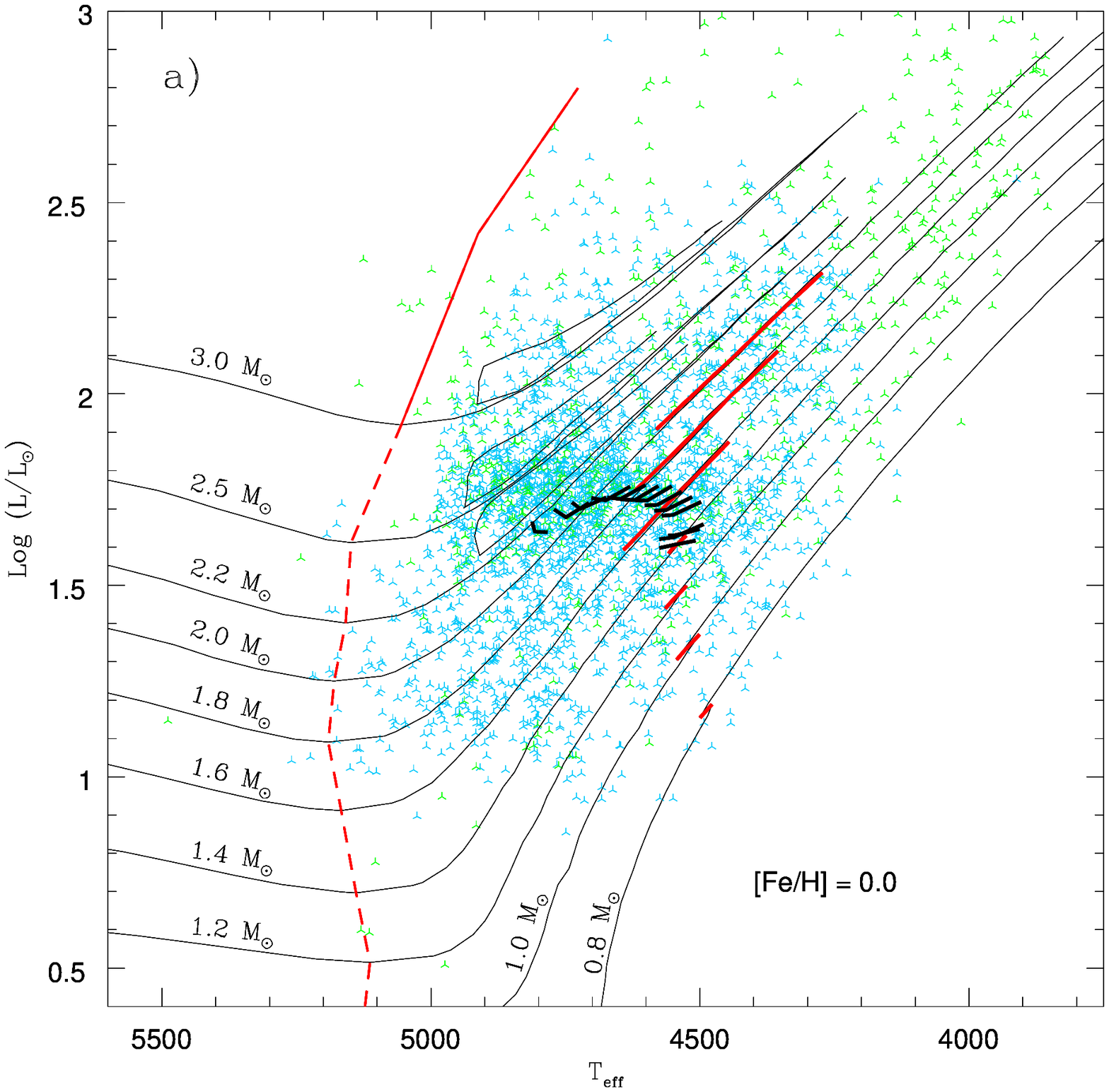}{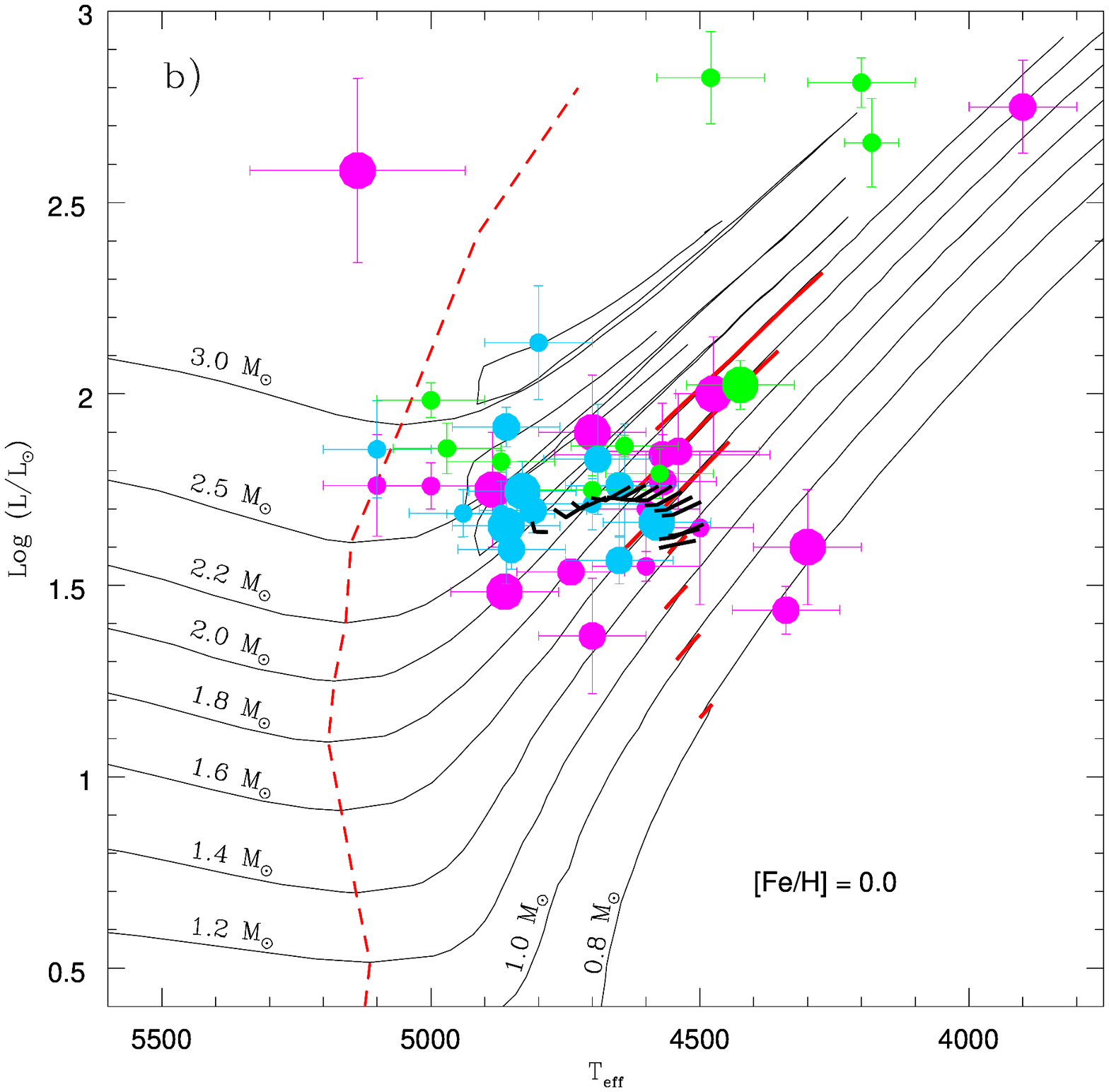}
\caption{Stars from our (blue symbols) and \citet{brown1989} (green
symbols) survey are shown (panel a) along with evolutionary tracks
computed by
\citet{bertelli2008}. Li-rich K giants are shown in
panel b: blue filled circles denote new Li-rich giants
found in this study ; green symbols are Li-rich giants from
\citet{brown1989}; magenta symbols are other Li-rich giants taken from the
literature.
Symbol size indicates
amount of Li.
The base of
the RGB is shown as a broken red line and
red portion on each of the tracks 
represents the location of the  luminosity bump which is 
predicted to be seen for masses $M \leq 2M_\odot$.
The thick black lines represent the clump region for He-core
burning stars of masses 0.8 - 2.5M$_{\odot}$.
\label{fig2}}
\end{figure}

All but a very few stars
in the two samples have evolved beyond the base of the
RGB (the red dashed line in Fig.~2)
and, therefore, their surface Li abundance has been greatly diluted.
In panel a of Fig.~2,
there is the expected concentration of stars with luminosities of
about
$\log \simeq 1.7$. This population in
the HR diagram in the main
are red clump (horizontal
branch) He-core burning stars that have evolved from
the tip of the RGB: the thick black line in Fig.~2 shows the
predicted locations of clump stars.
Extension of the $\log L/L_\odot \simeq 1.7$ concentration to
cooler temperatures may include stars at the bump luminosity of the first
ascent  RGB stars (see red portions of the evolutionary tracks).
Dominance of the HR diagram by clump stars is expected
because the lifetime of a clump star is typically one to two orders of
magnitude longer than for a bump star. This argument is not 
strictly applicable
to the Li-rich stars because their frequency is also dependent on the
probabilities  that clump and bump stars produce lithium.

Except for five luminous examples, the Li-rich stars are either
at the luminosity of the clump or within the range of expected values for
the bump, a range spanning the clump's luminosity.
The five luminous exceptions
must owe their Li richness to  different phenomena than those
accounting  for the majority of the Li-rich stars.
For example, HD~39853 \citep{gratton1989}  in Table 2 is a Li-rich K giant
but with log L/L$_{\odot}$=2.7
is probably a post-clump early AGB candidate where,
as discussed by \citet{charbonnel2000}, Li-production
by the Cameron-Fowler mechanism has occurred by a variant of its
operation in bump stars (see below).
In contrast, the high luminosity warm ($T_{\rm eff}=5150$K) Li-rich star HD 21018
is, as suggested
by \citet{charbonnel2000}, caught prior to dilution of lithium by the
convective
envelope, i.e., it is a Li-normal star exhibiting a Li abundance close to its
in initial abundance.
We focus here on the Li-rich
giants at bump and clump luminosities.

\section{Lithium-synthesis at the bump and/or clump?}

Since the Li-rich K giants
are concentrated within a narrow luminosity range, one may suppose a
significant
event must be associated directly or indirectly with this range.
One suspects that the significant event is either
the bump on the first ascent of the RGB and/or associated with the red clump formed
of He-core burning giants  that have experienced the He-core flash at the
tip of the RGB.

The observed luminosity range is crossed  by first ascent RGB stars  in which
the H-burning shell passes into the region that previously was the maximum
extent reached by the convective envelope.
When the outward moving H-burning shell, as a result of
increasing He-core mass,
first encounters this region, a
 star reacts to the increased availability of H by lowering
the surface luminosity. After a while, the giant continues its ascent of the
RGB. This creates a kink
on the evolutionary tracks for stars of masses
up to 2M$_{\odot}$
\citep{iben1968,bertelli2008} and a
bump in the luminosity
function
along the RGB.
As shown in Fig.~2, the kink on the evolutionary tracks \citep{bertelli2008}
is visible only up to 2M$_{\odot}$ for solar metallicity stars.

Internally, the kink is associated with an inversion in the
run of mean molecular weight with distance from the stellar center.
This inversion is linked to destruction of $^3$He by the reaction
$^3$He($^3$He,2$p$)$^4$He which
lowers the mean molecular weight, and the homogenization of the
composition within the convective enevelope. 
\citet{eggleton2008} show that this
inversion leads to `compulsory' mixing and changes to the
surface abundances of C,N, and O isotopic abundances, i.e., the
$^{12}$C/$^{13}$C
ratio is lowered relative to its value before the bump. \citet{charbonnel2010}
 recognize too that mixing occurs as a result of the
molecular weight
inversion but  include the effects of rotationally-induced mixing to drive
the mixing.
This mixing referred to as $\delta\mu$-mixing by \citet{eggleton2008} or
thermohaline
mixing by \citet{charbonnel2010} is observationally confirmed by
measurements of the
$^{12}$C/$^{13}$C ratio in giants along the RGB showing a decrease in the
ratio at and above the luminosity of the bump.

As $p$-captures on $^{12}$C create $^{13}$C,
the reservoir of primordial and
main-sequence synthesized  $^3$He is  depleted. It is this
reservoir
that is a potential source of $^7$Li from the Cameron-Fowler \citep{cameron1971}
mechanism ($^3$He($^4$He,$\gamma$)$^7$Be($e^-,\nu$)$^7$Li) but in order for
the
$^7$Li to enrich the stellar atmosphere it and its progenitor $^7$Be
must be
swept quickly to temperatures too cool for proton captures  to
occur.
\citet{eggleton2008} calculations show that more than about 80\% of the
$^3$He is destroyed in stars of masses less than about 1.5$M_\odot$. This
destruction seems unlikely to produce lithium because the mixing is too
slow for the $^7$Be and $^7$Li to avoid destruction by protons \citep{lattanzio2008}.
However, the initial  subsurface $^3$He reservoir is such that only a
minor fraction of the $^3$He need be converted with moderate
efficiency to provide a Li-rich giant. It is anticipated that the
lithium produced as the star
crosses the
bump's luminosity will be destroyed as the star with its convective
envelope evolves to the tip of the RGB.

The evidence from Fig.~2
is that few of the Li-rich stars are  aligned along the run of
bump stars in the HR diagram. Although the cooler Li-rich stars are likely
to be bump stars, many Li-rich stars are too warm to be
so identified.  Thus, we suggest that Li-rich
stars cannot be identified exclusively with the bump.
Rather the colocation of the Li-rich stars in Fig.~2 with the
high concentration of observed (Li-normal) giants and the
theoretical location of the clump suggests that they are
clump stars. Since Li, if produced at the bump on the RGB is almost certainly
destroyed
by the time the star has evolved up the RGB
 to experience the He-core flash, the
speculation is
that the Cameron-Fowler mechanism
operates at the He-core flash in at least some stars, i.e.,
in stars of $M < 2.25M_\odot$.
Since \citet{eggleton2008} predict survival of $^3$He in stars with $M>
1.5M_\odot$,
there seems in principal the possibility that the He-core flash  may be
the key to
synthesis of lithium in stars in a narrow mass range centered at
about 2$M_\odot$ with the range's the upper limit set by the maximum
mass for
a He-core flash and the lower limit set by survival of sufficient $^3$He
following
$\delta\mu$ or thermohaline mixing on the RGB.
Evidently, the concentration of Li-rich stars at the clump implies
that the synthesized Li is swiftly destroyed as a clump star evolves
along the early reaches of the AGB. If all stars evolving through the He-core
flash synthesize copious amounts of Li, they survive as a Li-rich
giant for about one per cent of their horizontal branch lifetime, i.e.,
about 2 Myr.
(The four luminous Li-rich stars in Fig.~2 have a mass in excess of those
experiencing the He-core flash. \citet{charbonnel2000} suggest
that these tap their $^3$He to produce Li as the He-burning shell 
evolves outward.)

\section{Conclusions}

The principal novel result here is that the majority of Li-rich
K giants have a luminosity and effective temperature combination
suggesting that Li production occurs at the He-core flash in those
stars. Although this speculation about the He-core flash has yet to
be supported by calculations, the nuclear physics of Li
production is surely that described by the Cameron-Fowler mechanism,
i.e., conversion of $^3$He to $^7$Li by $\alpha$-capture with $^7$Be as
a radioactive intermediary.
An earlier suggestion that Li production, also by the Cameron-Fowler
mechanism,  occurs at the
luminosity bump of the RIB \citep{charbonnel2000}
is required to account for cooler Li-rich stars.

Our discovery of that Li-rich giants are concentrated within a narrow
luminosity range does not support a view that Li-rich
giants result from the swallowing by the giant of a gaseous planet
\citep{alexander1967} or some other external origin. 

It remains to confirm and interpret suggestions that Li-rich K giants may
be   unusual with respect to other K giants in exhibiting rapid
rotation and/or an infrared excess (see, for example, \citealt{drake2002};
\citealt{charbonnel2000}). There are certainly Li-rich giants with
normal $^{12}$C/$^{13}$C ratios (e.g., HD 108471 in Table~2),
and/or rapidly rotating surfaces (e.g., HD 217352 with $v\sin i \simeq 35$ km s$^{-1}$,
\citealt{strassmeier2000}), and/or infrared excesses (e.g., PDS 365, \citealt{drake2002}). 

Perhaps, an area for observational scrutiny is a full determination
of the C, N, and O elemental and isotopic abundances in order to
search for discriminant between the candidate bump and clump Li-rich
stars themselves and between these Li-rich stars and bump and clump
stars exhibiting a normal Li abundance. Finally, a radial velocity
study should be undertaken to see if the collection of Li-rich stars have the
normal degree of binarity.

\acknowledgments
We are thankful to the trainees at CREST and VBO observatory staff for their help during observations, and
G. Pandey for his observant comments. 
DLL thanks the Robert A. Welch Foundation of Houston, Texas for support
through grant F-634. This research has made use of the
 SIMBAD database and the NASA ADS service.


\begin{table}
\caption{Derived results of the new Li-rich K giants.  
\label{tb1}}
\begin{tabular}{lrrrrrrrrrrccc}
\hline
Star & log(L/L$_{\odot}$) & T$_{\rm eff}$ & log$g$ & [Fe/H] & \multicolumn{2}{c}{log $\epsilon$(Li)} & $^{12}$C/$^{13}$C \\
& & & & &   $LTE$ & $NLTE$ &  \\
\hline
 HD~8676    &  1.68  &  4860  &   2.95   &     0.02   & 3.86 & 3.55 & 5.0 \\
 HD~10437   &  1.77  &  4830  &   2.85   &    0.10   &   3.76 & 3.48 & 5.0 \\
 HD~12203   &  1.69  &  4870  &   2.65   &  $-$0.27   &  2.01 & 2.08 &  7.5 \\ 
 HD~37719   &  1.76  &  4650  &   2.40   &   0.09    &  2.70 & 2.71 & \nodata \\ 
 HD~40168   &  2.10  &  4800  &   2.50   &    0.10   & 1.49 & 1.70 & \nodata \\ 
 HD~51367   &  1.59  &  4650  &   2.55   &   0.20     & 2.58 & 2.60 &  8.5 \\ 
 HD~77361   &  1.66  &  4580  &   2.35   &  $-$0.02   & 3.96 & 3.80 & 4.3 \\
 HD~88476   &  1.87  &  5100  &   3.10   &  $-$0.10   & 2.12 & 2.21 &  9.0 \\ 
 HD~107484  &  1.78  &  4640  &   2.50   &   0.18     &  2.04 & 2.14 & 12.5   \\
 HD~118319  &  1.68  &  4700  &   2.20   &  $-$0.25     &  1.88 & 2.02 & \nodata  \\ 
 HD~133086  &  1.70  &  4940  &   2.98  &     0.02   &  2.03 & 2.14 & 7.0 \\
 HD~145457  &  1.61  &  4850  &   2.75   &  $-$0.08   &  2.49 & 2.49 & 10.0 \\ 
 HD~150902  &  1.83  &  4690  &   2.55   &     0.09   &  2.64 & 2.65 &  5.0 \\
 HD~167304  &  1.93  &  4860  &   2.95   &    0.18   &  2.95 & 2.85 & 7.5 \\
 HD~170527  &  1.69  &  4810  &   2.85   &  $-$0.10   &  3.31 & 3.12 & \nodata \\
\hline
\end{tabular}
\end{table}

\begin{table}
\caption{Atmospheric parameters, Li abundance and carbon isotopic ratios of the known Li-rich giants.  
\label{tb2}}
\begin{tabular}{lrrrrrrrrrrccc}
\hline
Star & log(L/L$_{\odot}$) & T$_{\rm eff}$ & log$g$ & [Fe/H] & \multicolumn{2}{c}{log $\epsilon$(Li)} & $^{12}$C/$^{13}$C & Ref.\\
& & & & &   $LTE$ & $NLTE$ &  \\ 
\hline
HD~787       & 2.65$^{a}$ &     4181  & 1.5 & 0.07 &  1.80 & 1.99$^{a}$ & 9 & B00,Be94 \\
 HD~6665   &  1.37$^{a}$ &      4700$^{a}$ & 2.70$^{a}$ & 0.20$^{a}$ &  3.03$^{a}$ & 2.93$^{a}$ & \nodata & S00\\
 HD~9746    &  2.02$^{a}$  &  4425  &   2.30   &  $-$0.05   &  3.56$^{a}$ & 3.44$^{a}$  & 24 & B00 \\
 HD~19745        & 1.90   &  4700  &   2.25   &  $-$0.05   & 3.70 & 3.40 &  16  & R05\\
HD~21018$^{b}$     & 2.58$^{a}$ &     5150  &  1.96     &  0.15$^{a}$   &  3.13$^{a}$ &  3.06$^{a}$   & \nodata   &  B98   \\
HD~30834     & 2.81$^{a}$ &	  4200  & 1.5  & $-$0.17  & 1.80 & 1.98$^{a}$ & 13 & B00,Be94 \\
 HD~39853  &  2.75$^{a}$ &      3900 & 1.16 & $-$0.30 &  2.80$^{a}$  & 2.75$^{a}$ & 6.0  & G89\\
 HD~40827   & 1.78$^{a}$   &  4575  &   1.80   &    0.10   &  1.78$^{a}$ & 2.05$^{a}$  & 10$^{a}$  & B00 \\
 HD~63798   &  1.76$^{a}$  &  5000  &   2.50   &  $-$0.10   &  1.86$^{a}$ & 2.00$^{a}$  & 8.0$^{a}$ & M06 \\
 HDE~233517  &  2.00  &  4475  &   2.25   &  $-$0.37   &  4.11$^{a}$ & 3.95$^{a}$  & \nodata & Ba00\\
 HD~90633   &  1.55$^{a}$  &  4600  &   2.30   &     0.02    &  1.98$^{a}$ & 2.18$^{a}$  & 7.0$^{a}$ & M06 \\
HD~108471  & 1.86$^{a}$   &  4970  &   2.80   &  $-$0.01   &  1.96$^{a}$ & 2.10$^{a}$  & 25 & B00\\ 
 HD~112127  & 1.44$^{a}$   &  4340  &   2.10   &    0.09   &  3.01$^{a}$ & 2.95$^{a}$  & 19$^{a}$ & W82 \\
 HD~116292 &  1.82$^{a}$ &      5050 & 3.00 & $-$0.01 &  1.50  & 1.65 & \nodata & B00\\
PDS~365         & 1.85   &  4540  &   2.20   &  $-$0.09   & 3.30 & 3.13 &  12$^{+8}_{-2}$ & D02 \\  
HD~120602  & 1.98$^{a}$   &  5000  &   3.00   &  $-$0.08   &  1.95$^{a}$ & 2.07$^{a}$  & 16 & B00 \\
IRAS~13539-4153 & 1.60   &  4300  &   2.25   &  $-$0.13   & 4.10 & 3.90 &  20 & R05 \\
 HD~148293  & 1.86$^{a}$   &  4640  &   2.50   &    0.08   &  1.99$^{a}$ & 2.16$^{a}$  & 16  & B00\\
IRAS~17596-3952 & 1.70   &  4600  &   2.50   &    0.10   & 2.20 & 2.30 &  \nodata & R05\\
 HD~183492 &  1.75$^{a}$ &      4700 & 2.40 & $-$0.08 & 2.00$^{a}$ & 2.16$^{a}$ & 9 & B00 \\
PDS~100         & 1.65   &  4500  &   2.50   &    0.14   & 2.50 & 2.40 &  9.0 & R02 \\ 
HD~194937       & 1.54$^{a}$   &  4863  &   2.86   &  $-$0.01   & 3.41 & 3.18$^{a}$ & \nodata & L07\\
 HD~203136 &  1.75$^{a}$ &      5100$^{a}$ & 2.80$^{a}$ & 0.05$^{a}$ &  2.25$^{a}$ & 2.34$^{a}$ & \nodata & S00\\ 
HD~205349    & 2.82$^{a}$ &     4480  &  0.6  & 0.03    &  1.90 &  2.25$^{a}$   &  9   &  B00,Be94 \\
HD~214995  &  1.54$^{a}$  &  4740  &   2.56   &     0.00    &  3.16$^{a}$ & 2.95$^{a}$  & 13.0$^{a}$ & L07 \\
 HD~217352 &  1.77$^{a}$ &      4570 & 2.53 & \nodata$^{c}$ &  2.64 & 2.65$^{a}$ & \nodata & S00 \\
HD~219025       & 1.84$^{a}$   &  4570  &   2.30   &  $-$0.10   & 3.00 & 2.93$^{a}$ & \nodata & J99\\   
G0928+73.2600   & 1.75   &  4885  &   2.65   &  $-$0.25   & 3.62 & 3.30 &  28 & C10 \\
\hline
\end{tabular}
\tablenotetext{a}{From this work}
\tablenotetext{b}{Spectroscopic Binary and weak G-band star}
\tablenotetext{c}{Could not be measured due to large $vsini$ ($\simeq$ 35 km s$^{-1}$)}

\tablerefs{Ba00: \citet{balachandran2000}; B98: \citet{barrado1998}; Be94: \citet{berdyugina1994}; B00: \citet{brown1989};
C10: \citet{carlberg2010};
G89: \citet{gratton1989}; D02: \citet{drake2002}; J99: \citet{jasniewicz1999}; L07: \citet{luck2007}; M06: \citet{mishenina2006}; 
R02: \citet{reddy2002};  R05: \citet{reddy2005}; S00: \citet{strassmeier2000}; W82: \citet{wallerstein1982}  }
\end{table}

\end{document}